\pdfoutput=1

\documentclass[amssymb,amsmath,prl,twocolumn,superscriptaddress]{revtex4-1}

\usepackage[caption=false]{subfig}
\usepackage{graphicx}
\usepackage{bm}
\usepackage{amssymb,amsmath}
\usepackage{mathrsfs}
\usepackage{latexsym}
\usepackage{color}
\usepackage[normalem]{ulem}
\usepackage{dcolumn}
\usepackage[colorlinks,urlcolor=NavyBlue,citecolor=NavyBlue,linkcolor=NavyBlue,pdfusetitle]{hyperref}
\usepackage[usenames,dvipsnames]{xcolor}
\usepackage[utf8]{inputenc}
\usepackage{microtype}
\usepackage{etoolbox}
\usepackage{accents}

\allowdisplaybreaks

\definecolor{kcmagenta}{rgb}{0.54, 0.17, 0.88}
\definecolor{darkturquoise}{rgb}{0.0, 0.81, 0.82}

\newtoggle{commentsoff}
\togglefalse{commentsoff}

\ifdefined\nocomments
    \toggletrue{commentsoff}
\fi

\newtoggle{includeapp}
\toggletrue{includeapp}

\ifdefined\supplement
    \togglefalse{includeapp}
\fi

\iftoggle{commentsoff}{
  \newcommand{\kc}[1]{}
  \newcommand{\mi}[1]{}
  \newcommand{\wf}[1]{}

  \newcommand{\cn}[1]{}
  \newcommand{\del}[1]{}
}{
  \newcommand{\kc}[1]{{\textcolor{kcmagenta}{\sf{[KC: #1]}}}}
  \newcommand{\mi}[1]{{\textcolor{olive}{\sf{[MI: #1]}}}}
  \newcommand{\wf}[1]{{\textcolor{darkturquoise}{\sf{[WF: #1]}}}}

  \newcommand{\cn}[1]{{{\textcolor{blue}{\bf [cn]}}}}
  \newcommand{\del}[1]{\textcolor{red}{\sout{#1}}}
}

\graphicspath{{./fig/}}

\newcommand{\dcc}{LIGO--P1900109}

\newcommand{\p}{\hat{p}}
\newcommand{\deltap}{\delta \p}
\newcommand{\popmu}{\mu}
\newcommand{\popsig}{\sigma}
\newcommand{\eventmu}{\tilde{\mu}}
\newcommand{\eventsig}{\tilde{\sigma}}

\newcommand{\ppe}{\delta\hat{\varphi}}
\newcommand{\nevent}{N}
\newcommand{\nparam}{P}
\newcommand{\data}{\vec{d}}

\newcommand{\nn}{\nonumber}

\newcommand{\ts}{\textsuperscript}

\newcommand{\beq}{\begin{equation}}
\newcommand{\eeq}{\end{equation}}

\iftoggle{includeapp}{
\newcommand{\suppeq}{Eq.~\eqref{eq:inferred_distribution}}
}{
\newcommand{\suppeq}{Eq.~(5)}
}

\begin{document}

\title{Hierarchical test of general relativity with gravitational waves}

\newcommand{\CCA}{\affiliation{Center for Computational Astrophysics, Flatiron Institute, 162 5th Ave, New York, NY 10010}}
\newcommand{\MIT}{\affiliation{LIGO Laboratory and Kavli Institute for Astrophysics and Space Research, Massachusetts Institute of Technology, Cambridge, Massachusetts 02139, USA}}
\newcommand{\STBR}{\affiliation{Department of Physics and Astronomy, Stony Brook University, Stony Brook NY 11794, USA}}

\author{Maximiliano Isi}
\email{maxisi@mit.edu}
\thanks{NHFP Einstein fellow}
\CCA
\MIT

\author{Katerina Chatziioannou}
\email{kchatziioannou@flatironinstitute.org}
\CCA

\author{Will M. Farr}
\email{will.farr@stonybrook.edu}
\CCA
\STBR

\hypersetup{pdfauthor={Isi, Chatziioannou, Farr}}

\date{\today}

\begin{abstract}
We propose a hierarchical approach to testing general relativity with multiple gravitational wave detections.
Unlike existing strategies, our method does not assume that parameters quantifying deviations from general relativity are either common or completely unrelated across all sources.
We instead assume that these parameters follow some underlying distribution, which
we parametrize and constrain.
This can be then compared to the distribution expected from general relativity, i.e.~no deviation in any of the events.
We demonstrate that our method is robust to measurement uncertainties and can be applied to theories of gravity
where the parameters beyond general relativity are related to each other, as generally expected.
Our method contains the two extremes of common and unrelated parameters
as limiting cases.
We apply the hierarchical model to the population of $10$ binary black hole systems so far detected by LIGO and Virgo.
We do this for a parametrized test of gravitational wave generation, by modeling the population distribution of beyond-general-relativity parameters with a Gaussian distribution.
We compute the mean and the variance of the population and show that both are consistent with general relativity for all parameters we consider.
In the best case, we find that the population properties of the existing binary signals are consistent with general relativity at the ${\sim}1\%$ level.
This hierarchical approach subsumes and extends existing methodologies, and is more robust at revealing potential subtle deviations from general relativity with increasing number of detections.
\end{abstract}

\maketitle

\section{Introduction}

The ever-increasing number of binary coalescences~\cite{LIGOScientific:2018mvr} detected by LIGO \cite{TheLIGOScientific:2014jea} and Virgo \cite{TheVirgo:2014hva} has opened up avenues for rich new tests of general relativity (GR)~\cite{TheLIGOScientific:2016src,Yunes:2016jcc,TheLIGOScientific:2016pea,Abbott:2017vtc,Abbott:2018lct,LIGOScientific:2019fpa}.
This includes precision probes of strong-field orbital dynamics, the nature of the remnant object, and the properties of gravitational-wave (GW) propagation~\cite{TheLIGOScientific:2016src,LIGOScientific:2019fpa}.
With the new data, however, comes the problem of properly interpreting constraints in a way that does not apply only to specific modified theories of gravity and that is not biased by hidden assumptions~\cite{LIGOScientific:2019fpa,Zimmerman:2019wzo}.

In particular, there is an outstanding challenge to adequately combine information from different GW observations into a single statement about agreement with GR.
Existing approaches \cite{Li:2011cg,Li:2011vx,Agathos:2013upa,Meidam:2014jpa,DelPozzo:2011pg,Ghosh:2016qgn,TheLIGOScientific:2016pea,Meidam:2017dgf,Ghosh:2017gfp,Brito:2018rfr} rely on strong assumptions about the space of potential GR deviations and their effect on the observable events, rendering them too restrictive~\cite{Zimmerman:2019wzo}.
As a result, we might soon have a wealth of measurements from different techniques and events but no cohesive picture that brings them together.

In this paper, we present a flexible and robust solution to this problem by framing it in the language of hierarchical inference.
The result is an easy-to-interpret null-test of GR that can incorporate multiple measurements from different events,
without strong restrictions to specific theories of gravity or subclasses of events, and without the need to explicitly weigh events based on their significance.
We demonstrate that our method can produce strong combined constraints on deviations from GR.
If deviations are present, it can detect them even if they affect our measurements nontrivially, e.g.~by altering waveforms in ways that depend on the properties of each source.
We apply our method to GW detections from the GWTC--1 catalog of compact binaries~\cite{LIGOScientific:2018mvr,LIGOScientific:2019fpa}, using publicly available posterior samples for parameters controlling waveform deviations from the GR prediction \cite{GWOSC:O2TGR}.
We obtain joint constraints on deviations from GR that apply to generic theories of gravity, and find the data to be in agreement with Einstein's theory up to the ${\sim}1\%$ level.

\section{Method}

Waveform models for quasicircular compact binaries so far exist only within GR.
They are generically parametrized by $15$ parameters that describe the signal observed by an interferometric detector: component masses, component spins, location, orientation, and phase.
To date no parametrized waveform model exists that describes the inspiral, merger, and ringdown of generic binaries in any beyond-GR theory.
For this reason, and guided by the desire for model-idependent tests that do not
conform to specific theories, many studes are based on parametrized deviations away from the GR waveform.
In these tests, new degrees of freedom $\deltap_i$ are introduced,
 with $\deltap_i = 0$ corresponding to GR. These parameters are introduced to modify different aspects of
the waveform's frequency or amplitude evolution and, together with the $15$ GR parameters, define a generalized waveform model.
For more details on these parametric tests, see the Supplemental Material and~\cite{LIGOScientific:2019fpa} and references therein.

Unless they are somehow fixed to a constant by the true theory of gravity, we should generally
expect the $\deltap_i$'s to vary across different GW events. For
instance, the GW deformation could depend on the binary mass ratio or other
properties of the system, and different combinations of $\deltap_i$'s could come
into play under different circumstances. Without assuming a theory of
gravity, it is not possible to constrain the functional form of the
$\deltap_i$'s, making it difficult to combine measurements from different
events~\cite{LIGOScientific:2019fpa,Zimmerman:2019wzo}. To tackle this problem, we follow~\cite{Zimmerman:2019wzo} and employ a hierarchical formalism wherein we
assume that the true value of the beyond-GR parameters for each of the events is
drawn from some common unknown distribution \cite{2004AIPC..735..195L}. If
there are $\nparam$ parameters measured for $\nevent$ events, this amounts to
$\nparam\times\nevent$ random variables, which we denote $\deltap_i^{(j)}$ for
$i=1,\dots,\nparam$ and $j=1,\dots,\nevent$. Then each set of $\nevent$
variables corresponding to a given $\deltap_i$ should follow a shared
distribution, implicitly determined by the underlying theory of gravity and the
source population properties. The goal of the hierarchical approach (vividly
named ``extreme deconvolution'' by some \citep{Bovy:2009}) is to infer the
properties of the underlying distributions based on imperfect measurements from
a population of events.

The first step is to select a functional form for the distribution of $\deltap_i$, which is in principle nontrivial.
Given the small number of detections, here we only attempt to measure its mean $\popmu_{i}$ and standard deviation $\popsig_{i}$.
Higher moments, such as the skewness, could become measurable with an increasing number of detections.
In our case, and under a minimum-information assumption, we can thus model the population distribution with a Gaussian,
i.e.~we will take the population distribution to be $\deltap_i \sim \mathcal{N}(\popmu_i,\, \popsig_i)$.
A more complex function could be chosen as needed, with little impact on the method.
This potentially includes explicitly considering correlations among different $\deltap_i$'s, although we demonstrate below that this is not strictly necessary.

With the above choice of likelihood and appropriate values of $\popsig_i$, our method reduces to traditional nonhierarchical approaches for combining events~\cite{Zimmerman:2019wzo}.
Setting $\popsig_i =0 $ amounts to assuming that all systems share the same beyond-GR parameter
$\deltap^{(j)}_i = \popmu_i$.
The results are equivalent to multiplying the likelihood functions of the $\deltap^{(j)}_i$ for all detections $j$.
On the opposite extreme, letting $\popsig_i \rightarrow \infty$, the $\deltap^{(j)}_i $ are drawn from an effectively flat distribution and, as a result, measurement of one does not inform the others.
This corresponds to testing a theory of gravity in which each system is described by its own fundamental constant~\cite{Zimmerman:2019wzo}.
The results are equivalent to multiplying the Bayes factors from individual detections (assuming that a flat prior is imposed on each beyond-GR parameter).
However, both these assumptions can lead to incorrect conclusions if they do not apply to the true theory of gravity \cite{LIGOScientific:2019fpa,Zimmerman:2019wzo}.

In its general form, our hierarchical method is not limited by those assumptions and provides a robust way of detecting a deviation from GR even when the non-GR parameters are not trivially related to each other.
If GR is correct, then both hyperparameters, $\popmu_i$ and $\popsig_i$, are expected to be consistent with zero.
If we find a nonzero $\popmu_i$, this is an obvious deviation from GR or a systematic error in the analysis of one or several of the events under consideration.
Alternately, the true $\deltap^{(j)}_i$ could be symmetrically distributed around $\popmu_i=0$.
In this case, the inferred $\popmu_i$ will be consistent with $0$, but the $\popsig_i$ posterior will peak away from zero,
signaling that the scatter in $\deltap^{(j)}_i$ is larger than expected from statistical measurement errors, again revealing a beyond-GR effect (or modeling error).

\section{Simulation: GR is right}

\begin{figure}
\includegraphics[width=\columnwidth,clip=true]{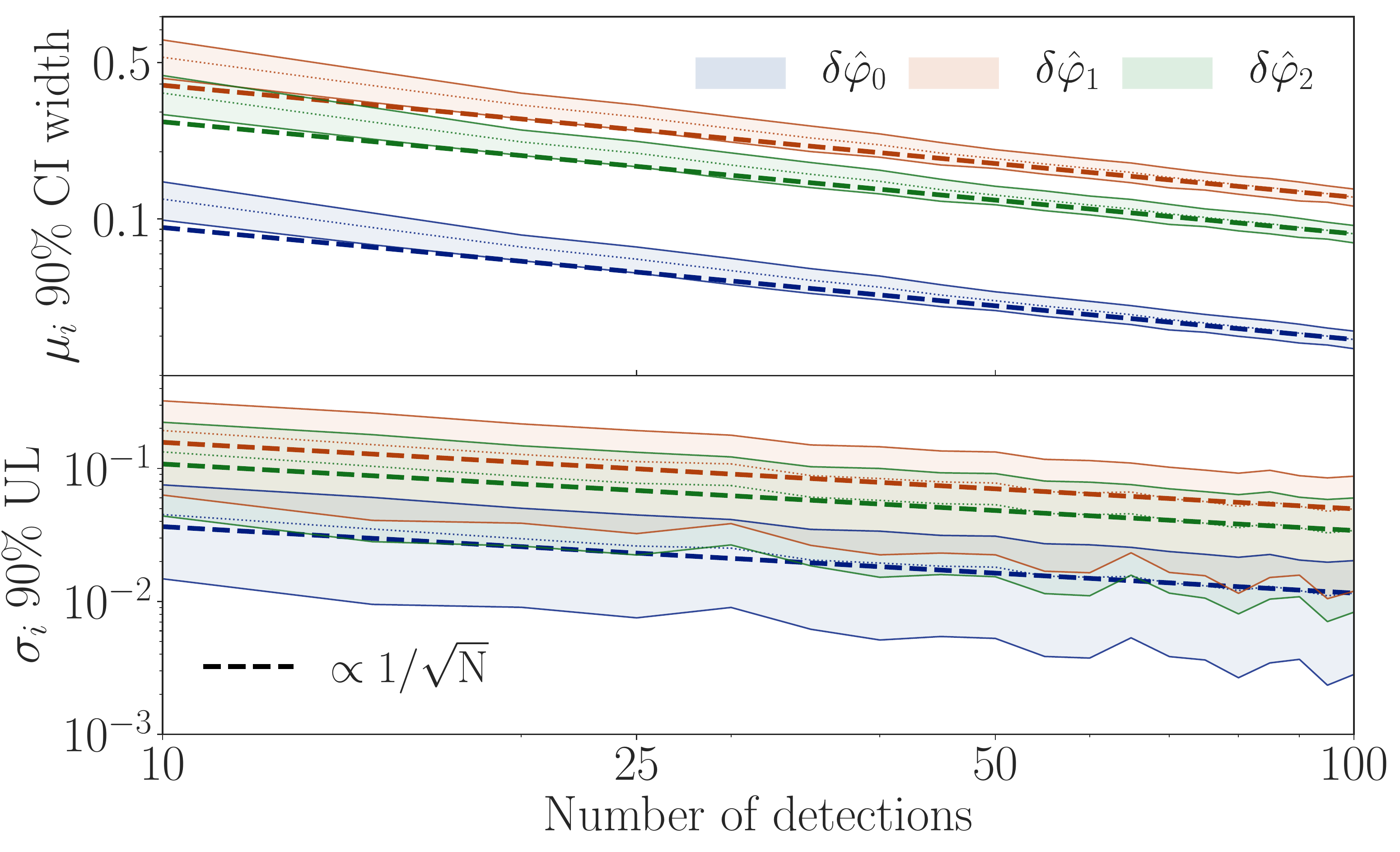}
\caption{Expected behavior of the population hyperparameters vs number of detections. We show the width of the $90\%$ credible interval for $\popmu_i$ (top) and the $90\%$ upper limit on $\popsig_i$ (bottom). In both panels
we average over 200 population realizations and shaded regions correspond to $1\sigma$
uncertainty. The dotted line show the mean over populations. The dashed line is proportional to $1/\sqrt{N}$; the bounds follow
the expected scaling with the number of detections.}
\label{fig:GRiscorrect}
\end{figure}

Given the long history of GR's experimental success~\cite{Will2014}, it is unavoidable to imagine that GW observations may also fail to reveal any shortcomings of the theory.
Accordingly, we begin by demonstrating our method on simulated signals that obey GR.
For simplicity, we take the measurement of each beyond-GR parameter to be summarized by a Gaussian likelihood with mean $ \eventmu^{(j)}_i$ and standard deviation $\eventsig^{(j)}_i$, i.e.~$p(\mathrm{data}^{(j)}\mid\deltap_i^{(j)}) = \mathcal{N}( \eventmu^{(j)}_i, \eventsig^{(j)}_i)$.
Such a likelihood is hardly realistic, especially for weak signals, but it suffices to illustrate our method and its scaling with the number of detections.
Note that $\eventmu^{(j)}_i$ and $\eventsig^{(j)}_i$ describe the idealized measurement of parameter $\deltap_i$ in the $j$\ts{th} event, while $\popmu_i$ and $\popsig_i$ define the distribution of true values of $\deltap_i$ across events.

We simulate a population of $\nevent$ observations as follows: first, we assign a random signal-to-noise ratio (SNR) to each event $j$ with the expected probability $\mathrm{SNR}^{(j)}\sim 1/\rm{SNR}^4$ \cite{2014arXiv1409.0522C}; then, for each $\deltap_i^{(j)}$, we assign a value of $\eventsig^{(j)}_i$ proportional to $1/\mathrm{SNR}^{(j)}$; finally, we choose a value of $\eventmu^{(j)}_i$ consistent with $\eventsig^{(j)}_i$ by drawing it from ${\cal{N}}(0,\eventsig^{(j)}_i)$, mimicking the expected scatter due to noise in the detector.
For concreteness, we consider only three non-GR parameters $\ppe_i$, $i=0,1,2$.
These are defined as in~\cite{LIGOScientific:2019fpa} and are related to the parametrized post-Einsteinian (ppE) framework of \cite{Yunes:2009ke}, as discussed in the supplement.
We set the overall scale of the $\eventsig^{(j)}_i$'s based on the uncertainty of measurements from GW150914 data, namely 68\%-level widths of 0.06, 0.3 and 0.2 for $\ppe_0$, $\ppe_1$ and $\ppe_2$ respectively~\cite{GWOSC:O2TGR}.

Figure~\ref{fig:GRiscorrect} shows the projected constraints on $\popmu_i$ (top) and $\popsig_i$ (bottom) for the ppE-like coefficients $\ppe_0$, $\ppe_1$ and $\ppe_2$ as the number of detections grows.
Colored bands represent the $1\sigma$ variation over 200 simulated populations.
The dashed line is proportional to $1/\sqrt{N}$ and demonstrates that bounds scale with the number of detections as expected.
Our method improves with increasing number of signals at a rate similar to the simple approach of multiplying the likelihoods, in spite of the presence of an additional parameter, $\popsig_i$.
This is because $\popmu_i$ and $\popsig_i$ are uncorrelated, so we can safely add $\popsig_i$
to our model without affecting the $1/\sqrt{N}$ scaling of $\popmu_i$, and vice versa.

\section{Simulation: GR is wrong}

\begin{figure}%
  \centering
  \subfloat{\includegraphics[width=\columnwidth,clip=true]{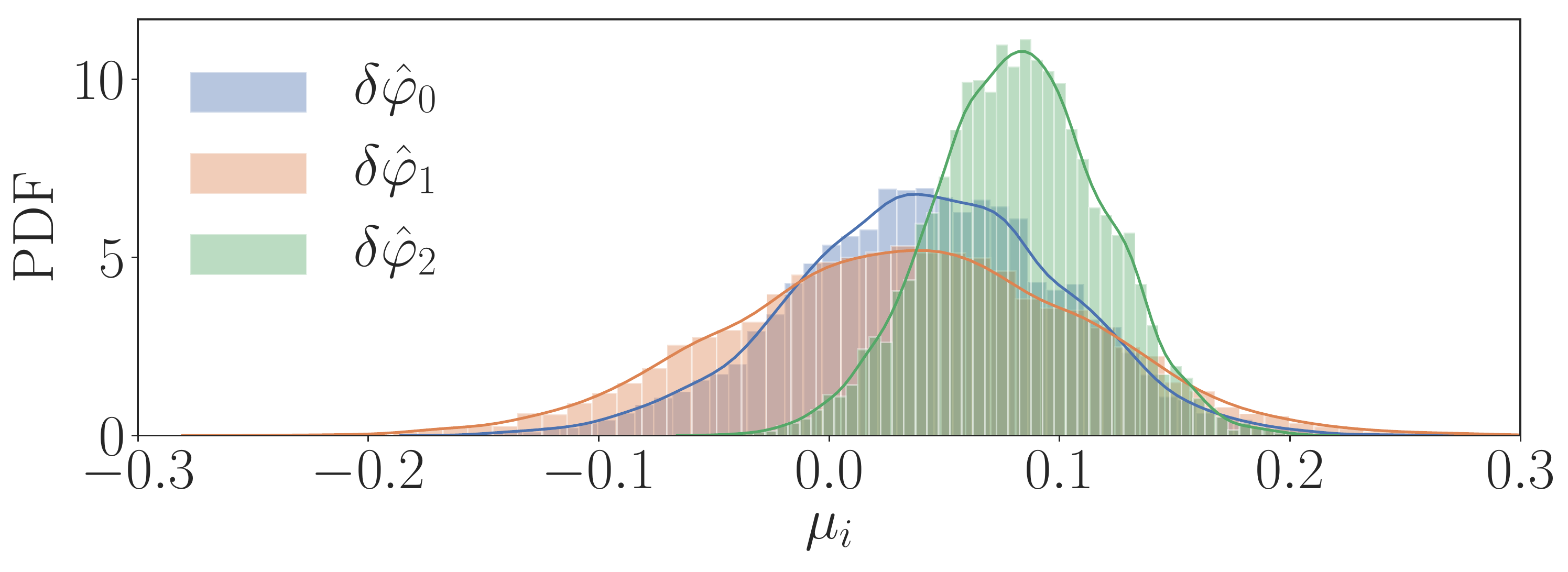}}\\
  \subfloat{\includegraphics[width=\columnwidth,clip=true]{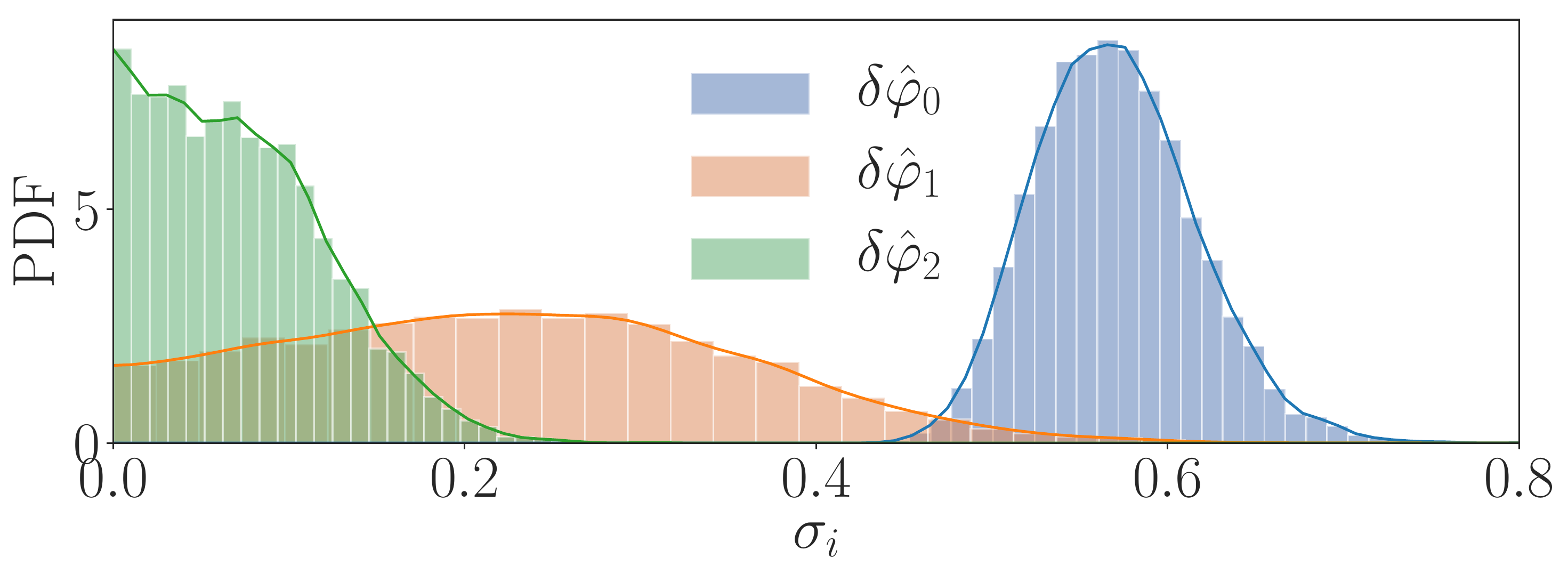}}
\caption{Example hyperparameter posteriors when GR is not the correct
theory of gravity. The deviation is only present at $\deltap^{(j)}_2=0.1$ but
it is recovered both in $\mu_2$ and $\sigma_0$. All other hyperparameters are consistent with GR.}
\label{fig:GRiswrong}
\end{figure}

We now turn to the tantalizing scenario that GR disagrees with experiment.
In such a case, we should generally expect the deviation from GR to manifest itself in multiple $\deltap_i$'s, even if it intrinsically occurs at a specific post-Newtonian (PN) order~\cite{Sampson:2013lpa,Meidam:2017dgf}.
This is because the phenomenological effect of modifications at different PN orders are not necessarily orthogonal, introducing degeneracies in our measurement.
Consequently, a deviation from GR affecting a given $\deltap_i$ could be measured through the $\popmu_i$ and $\popsig_i$ of multiple parameters, not just the one that is actually modified by the theory.

To demonstrate this effect, we construct a simple mock alternative theory of gravity that differs from GR at the 1PN order, affecting all binaries equally.
This intrinsic waveform correction is independent of source parameters, making it amenable to multiplication of the individual parameter likelihoods.
Generally, of course, this is not the case~\cite{Cornish:2011ys,Yunes:2016jcc}.
Even with this simplifying assumption, the \emph{measured} $\deltap_i$'s may vary in a nontrivial way with source properties
as signals with different frequency contents may be affected by the same deviation differently.

\begin{figure*}%
  \centering
  \subfloat{\includegraphics[width=0.333\textwidth,clip=true]{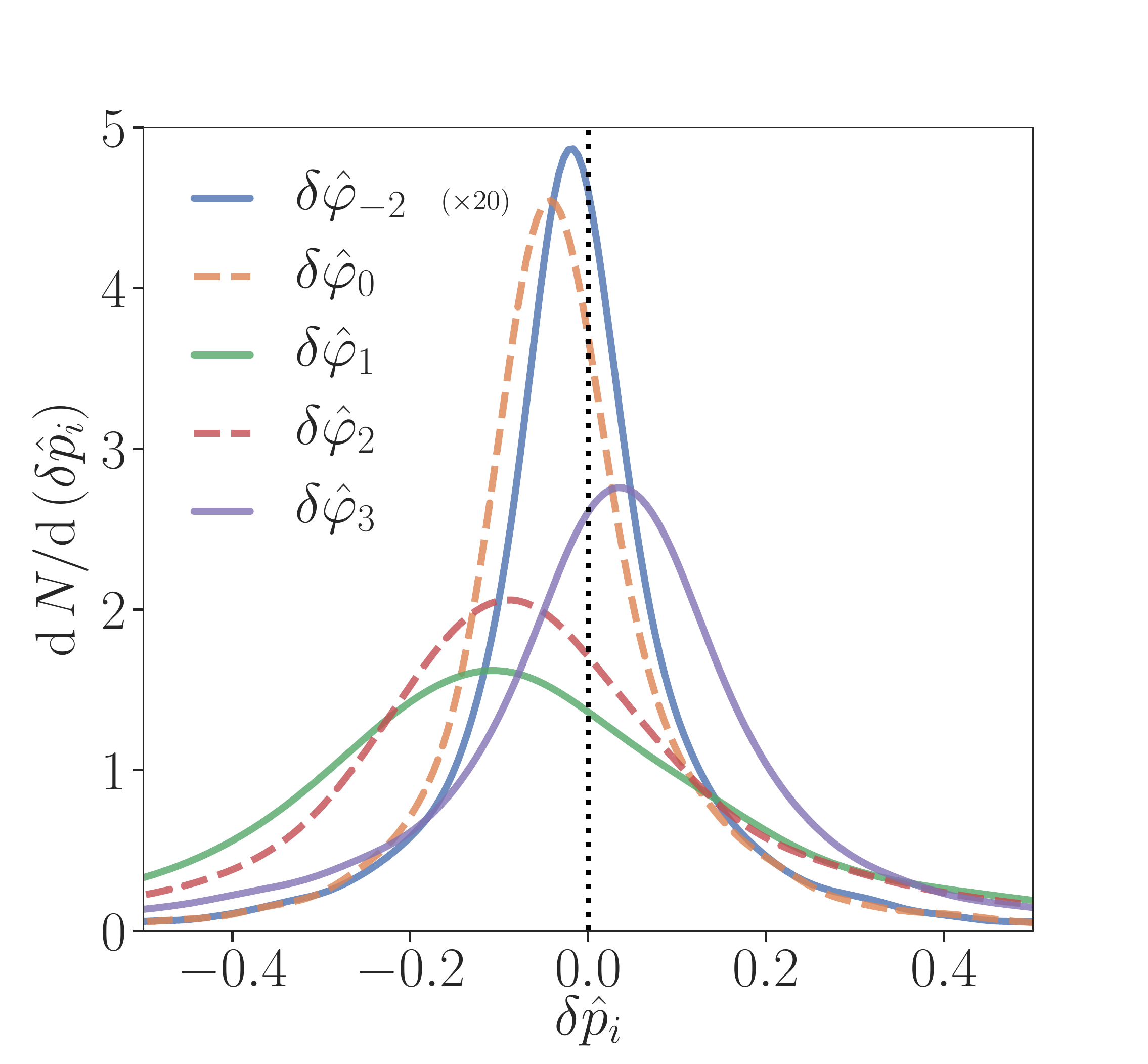}}\hfill
  \subfloat{\includegraphics[width=0.333\textwidth,clip=true]{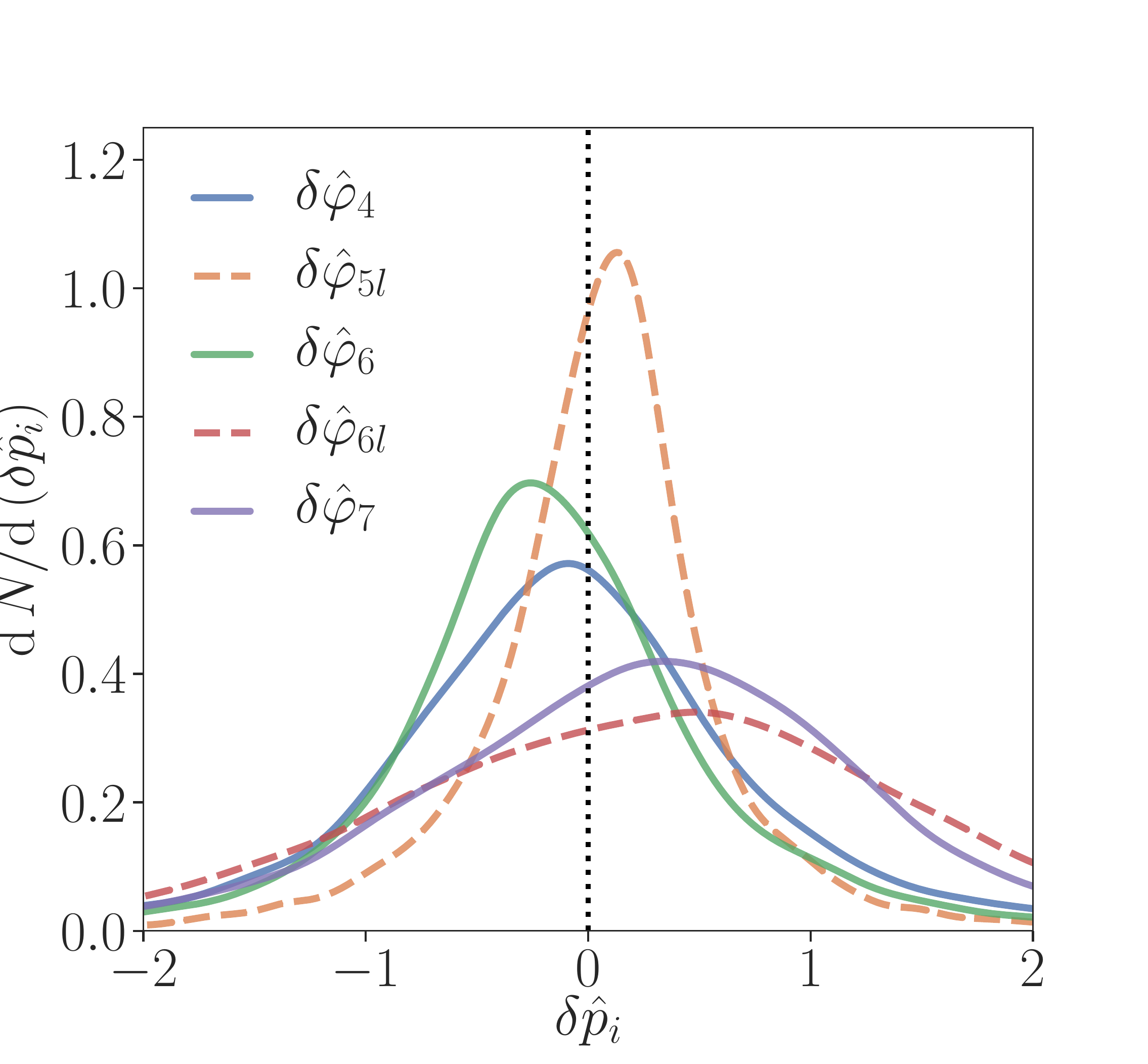}}\hfill
  \subfloat{\includegraphics[width=0.333\textwidth,clip=true]{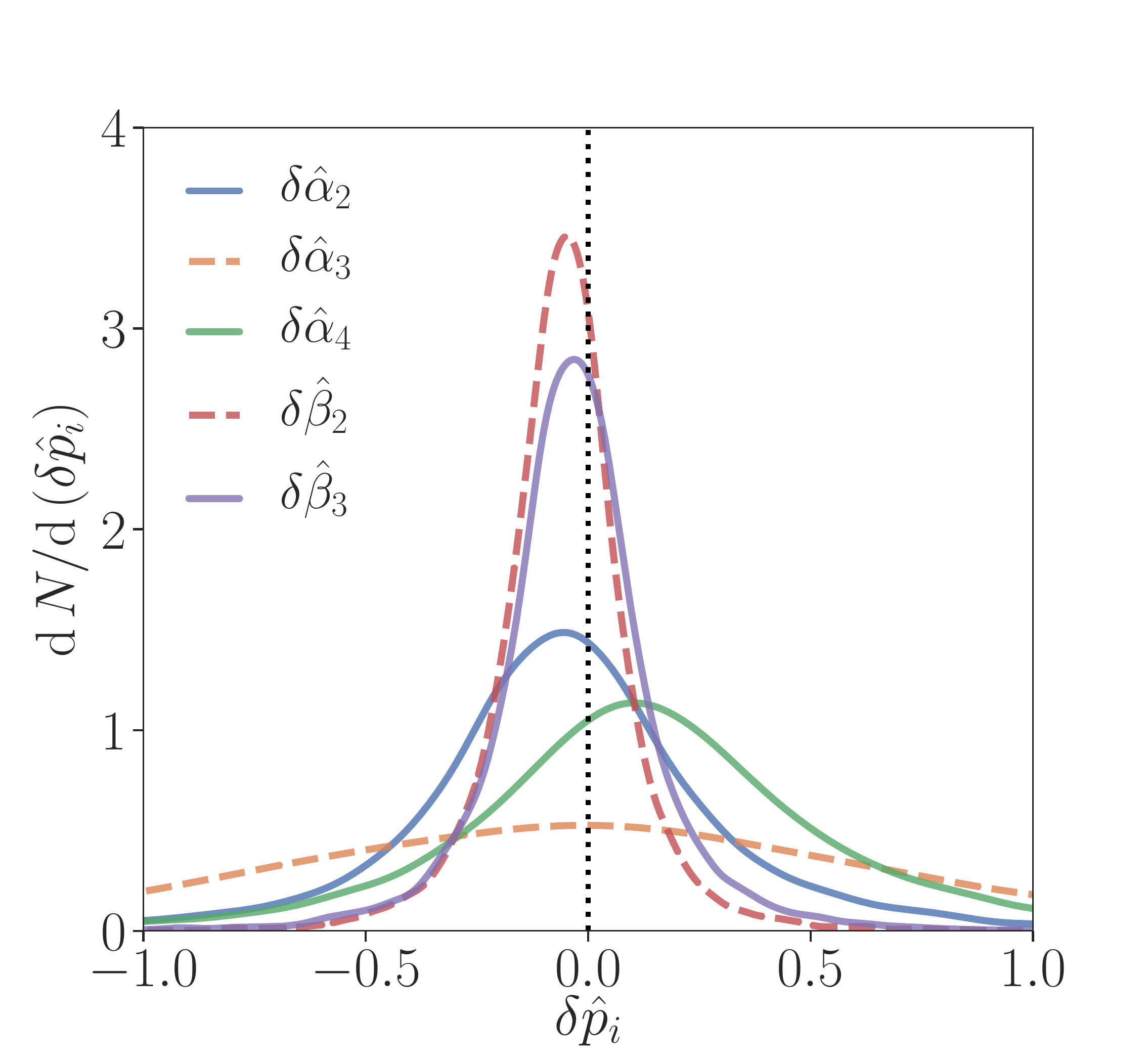}}
\caption{The inferred population distribution ${\rm d} N/ {\rm d} (\deltap_i)$ for the beyond-GR parameters $\deltap_i$'s given the $10$ confirmed binary black hole signals observed to date. (The scale of $\ppe_{-2}$ has been expanded by a factor of 20). All population distributions are consistent with $\deltap_i=0$, the GR prediction. With more observed signals, and under the assumption that they will obey GR, the population distributions are expected to become more narrowly centered around the origin, approximating a $\delta$-function.
These distributions are defined in \suppeq{} of the supplement.}
\label{fig:RealEvents}
\end{figure*}

Following~\cite{Meidam:2017dgf}, we assume that the \emph{measured} non-GR parameters $ \deltap_i $ depend nontrivially on the true values $\deltap^{\text{true}}_i$.
Generally, such relation could always be expressed via some measurement matrix $M$, such that $\deltap_i = M \deltap^{\text{true}}_i$, where the components of $M$ could depend on the specific properties of each system.
For our example, we again consider the three ppE-like parameters $\deltap_i = (\ppe_0,\, \ppe_1,\, \ppe_2)$ and we imagine $\deltap^{\text{true}}_i = (0,\,0,\,0.1)$, i.e.~the only parameter in which the modified theory deviates from GR is $\ppe_2$.
As an illustration, we arbitrarily pick a matrix $M$ that yields $ \deltap_i = (1.1-2 q,\, 0,\, 0.1)$,
where $q$ is the mass ratio of the system.
This is inspired by the degeneracy between high and low-order PN corrections demonstrated in \cite{Meidam:2017dgf}.
Quantitative results will be highly dependent on the true measurement matrix, though we only wish to demonstrate the qualitative
effect here.

We simulate a population of observations by drawing $q$ uniformly from $[0.1,1]$, and using those values to produce the measured parameters $\deltap_i$.
To simulate the corresponding posteriors, we draw the event SNRs and add a scatter due to noise as in the previous section.
As a result of the nontrivial dependence on $q$, the resulting population of each $\deltap_i$ is not normally distributed.
In spite of this, we demonstrate that our simple Gaussian model can detect the deviation from GR.

Figure~\ref{fig:GRiswrong}
shows the posteriors for $\popmu_i$ and $\popsig_i$ for a population of 100 events.
As expected, we find that the posterior for $\mu_2$ peaks at the injected value of $0.1$ and excludes GR at the $96$\% credible level.
Additionally, we find that $\sigma_0$ is not consistent with GR at the ${\gtrsim}99.99\%$ credible level.
This means that the scatter in $\ppe^{(j)}_0$ is too large to be accounted for
by statistical noise.
Indeed, part of the scatter in $\ppe^{(j)}_0$ is caused by the deviation from GR.
This illustrates that, even if we did not take $\ppe_2$ into account, we would have detected this deviation from GR solely through the lower PN order coefficient.
Additionally, the $\sigma_0$ posterior is farther from GR than the $\mu_2$ one, suggesting that this deviation could be detected first with a lower PN-order parameter.
We emphasize that these results are illustrative only: the properties of the posteriors in a real analysis will depend on the nature of the true measurement matrix, which is generally unknown.

\section{Real Events}

We now apply our hierarchical model to the confident binary black hole detections presented in GWTC--1 \cite{LIGOScientific:2018mvr}.
As a starting point, we use posterior samples for all $\deltap_i$ parameters from~\cite{LIGOScientific:2019fpa,GWOSC:O2TGR},
obtained with the \textsc{IMRPhenomPv2} waveform model~\cite{Hannam:2013oca,Khan:2015jqa}.
This study did not perform both sets of tests on all detected signals, but rather imposed certain thresholds on the SNR of the signals to determine whether to look for deviations in the inspiral or postinspiral regime, or both.
As a result, $5$ signals where analyzed for inspiral deviations and $9$ for postinspiral ones.
See \cite{LIGOScientific:2019fpa} for details.

As demonstrated in the previous sections, for each ppE-like coefficient $\deltap_i$, we obtain a posterior distribution for the corresponding hyperparameters $\popmu_i$ and $\popsig_i$.
We find that the population of the analyzed BBHs is consistent with GR both in terms of $\popmu_i$ and
$\popsig_i$ for all beyond-GR parameters. All $\popmu_i$ posteriors are consistent with $0$
at the 0.5$\sigma$ level or better, while all $\popsig_i$ posteriors peak at $0$.
This is a novel test, sensitive to generic beyond-GR effects in a population of detections.
With more signals we expect this analysis to either result in improved bounds (Fig.~\ref{fig:GRiscorrect}), or to reveal deviations from the ensemble properties expected from binaries in GR (Fig.~\ref{fig:GRiswrong}).

From the hyperparameter posteriors, we also compute the inferred population distributions for the $\deltap_i$'s, formally defined in \suppeq{} of the supplemental material and plotted in Fig.~\ref{fig:RealEvents}.
These distributions, ${\rm d} N/ {\rm d} (\deltap_i)$, represent our best knowledge of the population from which the allowed deviations from GR, $\deltap_i^{(j)}$, were drawn for each binary black hole signal.
Because all of these distributions contain zero with high probability, their width indicates the level to which we can constrain our measurements of the ppE-like coefficients to agree with GR.
In the best case, for $\ppe_{-2}$, we find consistency with GR at the ${\sim}1\%$ level with $68\%$ credibility.
In the future, and assuming GR is correct, we should find that the distributions tend to a $\delta$-function at $\deltap_i=0$ as we accumulate more observations (cf.~Fig.~\ref{fig:GRiscorrect}).
We emphasize that, unlike Fig.~3 in~\cite{LIGOScientific:2019fpa}, Fig.~\ref{fig:RealEvents} here does not show the inferred posterior on the ppE-like parameters assuming all signals share the same value.
Instead, Fig.~\ref{fig:RealEvents} summarizes our inference for the distributions from which the potentially unequal ppE-like parameters of each signal were drawn
\footnote{This is not the same as the most likely such distribution, which by construction would be a Gaussian with mean and variance given by the peak of the posterior on $\popmu_i$ and $\popsig_i$. Rather, Fig.~\ref{fig:RealEvents} \emph{marginalizes} over $\popmu_i$ and $\popsig_i$, which is why the ${\rm d} N/ {\rm d} (\deltap_i)$ distributions in Fig.~\ref{fig:RealEvents} are not Gaussians. See supplement for details.}.

These results are subject to the thresholds imposed in~\cite{LIGOScientific:2019fpa} that determine which GW events are subject to each test.
They would thus be vulnerable to the same potential selection effects.
This includes the requirement that signals be sufficiently loud and akin to GR, such that they are detectable by matched-filtering procedures looking for GR signals.
Reference~\cite{LIGOScientific:2019fpa} argues that both types of potential selection effects are partially mitigated by the fact that more generic searches are also employed alongside matched-filter ones.
With this caveat in mind, we find no evidence of any deviation from GR.

\section{Conclusions}

We use a hierarchical approach to test GR with GWs by assuming that beyond-GR parameters in each event are drawn from a common underlying distribution.
This approach is flexible and powerful, as it can encompass generic population distributions even if the chosen parametrization is inaccurate.
It can trivially incorporate future detections and  can be applied to different kinds of tests of GR, including searches for modified dispersion relations~\cite{Mirshekari:2011yq,Abbott:2017vtc} or inspiral-merger-ringdown consistency checks~\cite{Ghosh:2016qgn,Ghosh:2017gfp}.
We apply this method to the current 10 confident binary black hole detections~\cite{LIGOScientific:2018mvr}, measuring posterior distributions for the mean and standard deviation of the population of ppE-like parameters $\deltap_i$~\cite{GWOSC:O2TGR}.
This is a conceptually new test that examines the ensemble properties of GW signals rather than the properties of individual events; we find the set of measurements to be consistent with GR (Fig.~\ref{fig:RealEvents}).

Parametrized tests, such as the ones studied here, are powerful probes of beyond-GR effects.
Yet, it has long been appreciated that their interpretation demands caution: correlations between parameters
make it necessary to have a consistent model to characterize a detected deviation.
Our method provides a framework to execute a null test of GR with several detections, largely without the need for specific models of potential deviations.
It improves with increasing number of signals at a rate similar to simpler approaches (Fig.~\ref{fig:GRiscorrect}).
Furthermore, hierarchical methods could exploit degeneracies in our measurements to detect otherwise inaccessible deviations from GR, e.g.~because they intrinsically occur at a higher PN order than can be directly probed (Fig.~\ref{fig:GRiswrong}).

The framework presented here is not restricted to tests of GR with GWs, but can be generalized to include
information from other observations. For example, the measured likelihood for $\ppe_{-2}$ from GWs
could be combined with corresponding constraints from binary pulsar measurements.
Our hierarchical method not only unifies the signals seen by ground-based detectors, but also offers a way to consider multiple tests of GR simultaneously.

\begin{acknowledgments}
\section{Acknowledgments}

We thank Aaron Zimmerman and Carl-Johan Haster for useful discussions.
We thank Nathan Johnson-McDaniel for comments on the draft.
Samples from the $\popmu_i$ and $\popsig_i$ posteriors were drawn with {\tt stan}~\cite{JSSv076i01}, and
plots were produced with {\tt matplotlib}~\cite{Hunter:2007}.
M.I.~is supported by NASA through the NASA Hubble Fellowship grant \#HST--HF2--51410.001--A awarded by the Space Telescope Science Institute, which is operated by the Association of Universities for Research in Astronomy, Inc., for NASA, under contract NAS5--26555.
The Flatiron Institute is supported by the Simons Foundation.
This paper carries LIGO document number \dcc{}.
\end{acknowledgments}

\bibliography{OurRefs}

\appendix

\iftoggle{includeapp}{
    \newcommand{\includeapp}{}
    \section{APPENDIX: TECHNICAL DETAILS}
    
\ifdefined\includeapp
    \newcommand{\FigRealEvents}{Fig.~\ref{fig:RealEvents}}
\else
    \newcommand{\FigRealEvents}{Fig.~3}
\fi

\section{Hierarchical inference}

Our goal is to estimate the \emph{posterior} probability density representing the measurement of some parameters of interest, 
conditional on some data or observation.
In order to do this, it is necessary to write down a \emph{prior} probability density function that encodes 
knowledge about the possible values of the parameters before obtaining the data.
We use Bayes' theorem to obtain the posterior from the \emph{likelihood}, which is the probability of the data conditional on the values of the parameters.
For a parameter $x$ and set of data $\vec{d}$, Bayes' theorem is just
\beq
p(x \mid \vec{d}) \propto p(\vec{d} \mid x)\, p(x)\, ,
\eeq
relating the posterior (left) to the likelihood times the prior (right), with a proportionality constant that ensures unitarity.

The specific functional form of the prior is determined according to different expectations for the parameters.
For example, in situations of high ignorance about the expected values, common choices for the prior are a uniform distribution across some broad range 
or a least-informative (aka, ``Jeffrey's'') prior.
Alternatively, the prior may take the form of a multivariate normal distribution with some mean vector and covariance matrix.
The mean informs the analysis of the likely location of the parameter values in the $\nparam$-dimensional parameter space, 
and the covariance sets the scale of the corresponding uncertainty.
Just as with the boundaries of the flat prior, the mean and covariance matrix can be set arbitrarily.

An extension to the above fixed priors is the case where we allow this distribution to vary in some controlled way.
One such example is the case where we wish to marginalize over assumptions about the provenance of a parameter.
Alternatively, as in the main text, besides the values of the parameters for any given observation, we might also be 
interested in the distribution of said parameter in a population of observed data.
Both types of analyses make use of \emph{hierarchical models}, with multiple nested levels of inference with corresponding priors.
Priors on parameters that control distributions of other parameters (e.g. the mean and the covariance of the example priors above) 
are often called \emph{hyperpriors}.

In this paper, we perform two inference levels: the first corresponds to the measurement of ppE-like parameters from individual events, and the second is the measurement of the properties of the population of ppE-like parameters.
Concretely, we begin from the posteriors on $\nparam$ parameters $\deltap_i^{(j)}$ from the data $\data^{(j)}$ for the $\nevent$ events.
The posterior obtained from event $j$ for parameter $i$ takes the form:
\begin{align} \label{eq:single_posterior}
p(\deltap_i^{(j)} \mid \data^{(j)}) &\propto p( \data^{(j)} \mid \deltap_i^{(j)})\, p(\deltap_i^{(j)})\nn \\
&\propto p( \data^{(j)} \mid \deltap_i^{(j)})\, ,
\end{align}
where the last line is only valid under the assumption of a uniform prior on $\deltap_i^{(j)}$, as was applied in the original measurements by LIGO and Virgo \cite{LIGOScientific:2019fpa,GWOSC:O2TGR}.

We then assume that the values for each event are drawn from a normal distribution with mean $\popmu_i$ and standard deviation $\popsig_i$, characteristic of each parameter.
Since the goal is to measure those quantities from the data from \emph{all} events, for each $i$ we want the posterior
\begin{align} \label{eq:population_posterior}
p(\popmu_i,\popsig_i \mid \{\data^{(j)}\}_{j=1}^\nevent) &\propto p(\{\data^{(j)}\}_{j=1}^\nevent \mid \popmu_i,\popsig_i)\, p(\popmu_i,\popsig_i)\nn\\
&\propto p(\popmu_i,\popsig_i) \prod_{j=1}^\nevent p(\data^{(j)}\mid \popmu_i,\popsig_i)\, ,
\end{align}
where we took advantage of the fact that events are statistically independent to factorize the likelihood in the last line.
Each of those factors may be written explicitly in terms of the parameters $\deltap_i^{(j)}$, 
\begin{align} \label{eq:posterior_integral}
p(\data^{(j)}\mid \popmu_i,\popsig_i) = \int &p(\data^{(j)}\mid \deltap_i^{(j)})\times\nn\\
&p(\deltap_i^{(j)}\mid \popmu_i,\popsig_i)\, {\rm d}[\deltap_i^{(j)}]\, ,
\end{align}
obtaining an expression in terms of the individual likelihoods, Eq.~\eqref{eq:single_posterior}.
These likelihoods, $p(\data^{(j)}\mid \deltap_i^{(j)})$---or, rather, the corresponding posteriors with a suitable choice of prior---are what is computed in a regular (nonhierarchical) analysis with the goal of measuring $\deltap_i^{(j)}$,~as is the case in \cite{LIGOScientific:2019fpa}.

The last two equations allow us, then, to measure the population mean and standard deviation for the ppE-like parameters starting from the posterior samples released by the LIGO and Virgo collaborations \cite{GWOSC:O2TGR}, without re-analyzing the raw data (\FigRealEvents{}).
Specifically, we obtain posteriors on the population parameters, Eq.~\eqref{eq:population_posterior}, by sampling over $\popmu_i$ and $\popsig_i$ with a Hamiltonian Monte Carlo algorithm implemented in the \texttt{stan} package~\cite{JSSv076i01}.
To simplify the integration over the $\deltap_i$'s in Eq.~\eqref{eq:posterior_integral}, we internally represent the LIGO-Virgo posteriors via a Gaussian-kernel density fit produced from the samples in \cite{GWOSC:O2TGR}.

From a posterior on the hyperparameters, $p(\popmu_i,\popsig_i \mid \{\data^{(j)}\}_{j=1}^\nevent)$, we may also infer the shape of the population distributions themselves.
This can be done by marginalizing over $\popmu_i$ and $\popsig_i$,
\begin{align} \label{eq:inferred_distribution}
p(\deltap_i \mid \{\data^{(j)}\}_{j=1}^\nevent) = \int &p(\deltap_i \mid \popmu_i, \popsig_i)\, \times \nn\\
&p(\popmu_i,\popsig_i \mid \{\data^{(j)}\}_{j=1}^\nevent)\, {\rm d} \popmu_i {\rm d}\popsig_i\, ,
\end{align}
where $p(\ppe_i \mid \popmu_i, \popsig_i) = {\cal N}(\popmu_i, \popsig_i)$ by construction.
In the main text we label the inferred distribution of Eq.~\eqref{eq:inferred_distribution} as ``${\rm d}N/{\rm d}(\deltap_i)$,'' since it can be interpreted as the expected fractional number of events ${\rm d}N$ with a value of the ppE-like coefficient $\deltap_i^{(j)}$ within $[\deltap_i, \deltap_i+{\rm d}(\deltap_i)]$ .

\section{Parametrized tests of GR}

In this study, we consider parametric deformations to the GW signal described by some non-GR quantities 
$\deltap_i$ indexed by $i$, with $\deltap_i = 0$ corresponding to GR.
These new parameters are introduced on top of the 15 usual parameters that describe a GW within GR, with the goal of providing a 
model-independent framework with which to test the theory. The parameters are generally chosen such that they modify a specific aspect 
of the waveform, and hence a non zero value would signal a specific type of violation of GR.
The parametric tests we will consider here are introduced to test the waveform in the different regimes of a binary coalescence: the inspiral, the merger, and the ringdown. 

The inspiral phase is deformed through the post-Einsteinian (ppE) inspiral parameters $\delta\p_i=\ppe_i$~\cite{Yunes:2009ke}.
We define these $\ppe_i$'s as in~\cite{LIGOScientific:2019fpa}, a choice that differs slightly from that of~\cite{Yunes:2009ke} but which is analytically equivalent~\cite{Yunes:2016jcc}.
The usual inspiral phase within GR is usually expressed as an expansion in small velocities,
 referred to as the post-Newtonian (PN) expansion. Each
coefficient in the expansion depends on the system parameters and measurement of the GW phase evolution amounts to measuring said parameters.
The ppE parameters are introduced as relative (or absolute, in some cases where the GR term vanishes) deformations of the normal PN coefficients such that 
$\ppe_i$'s encodes a correction at the $i/2$ PN order. For example, $\ppe_{-2}$ corresponds to a -1PN correction, associated with dipole radiation.

The merger-ringdown and the intermediate regimes are deformed through a different set of post-inspiral parameters, denoted $\delta\hat{\alpha}_i$ and $\delta\hat{\beta}_i$ respectively~\cite{Meidam:2017dgf}.
These parameters encode modifications to the analytic description of those post-inspiral stages as implemented in the \textsc{IMRPhenomPv2} waveform model~\cite{Hannam:2013oca,Khan:2015jqa}.
In particular, the $\delta\hat\alpha_i$'s control the merger-ringdown coefficients $\alpha_i$, which are obtained both from phenomenological fits and black-hole perturbation theory~\cite{Khan:2015jqa}.
The $\delta\hat\beta_i$'s control deviations from the NR-calibrated phenomenological coefficients
$\beta_i$ of the intermediate stage.

\ifdefined\includeapp
\else
    \bibliography{OurRefs}
    \end{document}
\fi

}{
}

\end{document}